\documentclass[copyright]{eptcs}

\providecommand{\event}{MARS 2020}

\usepackage{underscore}
\usepackage{url}
\usepackage{breakurl}
\usepackage{xspace}
\usepackage{graphicx}
\usepackage{subcaption}

\usepackage{listings}
\usepackage{cadp-lotos}
\usepackage{cadp-lnt}

\usepackage{xcolor}

\definecolor{grey}{cmyk}{0,0,0,0.8}
\definecolor{gray}{cmyk}{0,0,0,0.5}
\hypersetup{bookmarks=true,bookmarksopen=false,colorlinks=true,linkcolor=grey,citecolor=grey,urlcolor=grey}

\newcommand{\I}[1]{\mbox{\em #1\/}}

\newcommand{\T}[1]{\mbox{\tt #1}}

\newcommand{\TLA}{TLA+\xspace}

\newcommand{\IGNORE}[1]{}

\title{Modeling the Raft Distributed Consensus Protocol in LNT}

\def\titlerunning{Modeling the Raft Distributed Consensus Protocol in LNT}

\author{Hugues Evrard
   \institute{Google}
   \email{hevrard@google.com}
}

\def\authorrunning{H.~Evrard}

\begin{document}

\maketitle

\begin{abstract}
Consensus protocols are crucial for reliable distributed systems as they let them cope with network and server failures.
For decades, most consensus protocols have been designed as variations of the seminal Paxos, yet in 2014 Raft was presented as a new, ``understandable'' protocol, meant to be easier to implement than the notoriously subtle Paxos family.
Raft has since been used in various industrial projects, e.g. Hashicorp's Consul or etcd (used by Google's Kubernetes).
The correctness of Raft is established via a manual proof, based on a \TLA specification of the protocol.
This paper reports our experience in modeling Raft in the LNT process algebra.
We found a couple of issues with the original \TLA specification of Raft, which has been corrected since.
More generally, this exercise offers a great opportunity to discuss how to best use the features of the LNT formal language and the associated CADP verification toolbox to model distributed protocols, including network and server failures.
\end{abstract}



\section{Introduction}

Consensus protocols enable distributed systems to cope with network and server failures via the state machine replication approach \cite{Schneider-90}.
Most consensus protocols are designed as variations of Paxos \cite{Lamport-98}, and they are all notoriously difficult to implement due not only to their inherent  complexity, but also to the fact that they are typically presented in abstractions that are non-trivial to transcribe into an executable program.
The Raft consensus protocol was designed with understandability in mind, and user studies indicate that it is easier to reason about than Paxos \cite{Ongaro-Ousterhout-14}.
Several open-source implementations of Raft exist,\footnote{\url{https://raft.github.io/\#implementations}} and the protocol is also used in some industrial projects like Hashicorp's Consul\footnote{\url{www.consul.io/docs/internals/consensus.html}} or the etcd\footnote{\url{github.com/etcd-io/etcd/tree/master/raft}} key-value store used by Google's Kubernetes.\footnote{\url{kubernetes.io/blog/2019/08/30/announcing-etcd-3-4/}}
Raft is formally defined by its \TLA specification \cite{Ongaro-14}, upon which the Raft authors presented a hand-written proof of the protocol safety.

In this paper, we discuss our experience in modeling Raft in the LNT formal language \cite{Champelovier-Clerc-Garavel-et-al-10-v6.7}.
A first version of this model was written as a use-case for the Distributed LNT Compiler \cite{Evrard-16}, and helped to identify two issues in the original \TLA specification, which have been fixed since.
We contributed a second version as a new model for the 2015 edition of the Model Checking Contest \cite{mcc:2015-results,mcc2015}.
For the first time, in this paper, we have the opportunity to discuss the whole specification in detail.
Our goal is not so much to describe Raft internals as to discuss the process of developing the formal model of a distributed system using LNT.

An LNT model can be verified using CADP \cite{Garavel-Lang-Mateescu-Serwe-13}, which relies on explicit state-space exploration model checking techniques.
The style in which an LNT model is written can have a dramatic effect on the size of its state space, possibly restricting the scope of achievable verification.
Hence, we regularly mention which style choices were made with respect to their impact on the state space size.

The paper is structured as follows:
Section~\ref{OVERVIEW} gives an overview of Raft, and Section~\ref{MODEL} presents its LNT model.
Section~\ref{DISCUSSION} discusses the style of the specification, and broadens to distributed systems modeling in general.
Section~\ref{CONCLUSION} gives concluding remarks.
The complete LNT specification of Raft is provided in Annex~\ref{SPECIFICATION}.


\section{The Raft consensus protocol}\label{OVERVIEW}

We give a brief overview of the consensus problem, and how Raft solves it.
We aim at providing a general understanding of Raft structure, more details come in the following section where we present the LNT model.


\subsection{Consensus for state machine replication}

The state machine replication approach \cite{Schneider-90} enables a service to be made robust to failures by replicating the logic of the service, represented by a deterministic state machine, among several servers of a distributed system.
When a service's client request---named \I{command} here---arrives, a server handles it.
This server then replicates them to all servers, which store them into their local \I{log}.
As long as all server logs stay coherent, they enable servers process the same client commands in the same order, such that all the replicated deterministic state machines reach the same states and return the same responses.
Consensus is the mean by which we keep server logs coherent.
This lets the system cope with server failures: the service is kept available via the servers that are still alive.

This approach relies on a consensus protocol whose role is to let all alive servers agree on which client commands to add to their log, and in which order, and to do so even in the presence of server and network failures.
Raft is a consensus protocol that operates in two main phases: leader election, and log replication.


\subsection{Raft leader election}

In Raft, time is divided into \I{terms}, which are monotonically increasing and act as
logical clocks \cite{Lamport-78}.
The leader election phase of Raft aims at getting one server to be elected as the leader for the current term.

Servers can be in one of three states: \I{follower}, \I{candidate} or \I{leader}.
Every server that is not a leader has a timeout after which, if it did not receive any message, it increments its own term index, moves into the candidate state, and sends vote requests to all other servers.
A server that receives a vote request will grant its vote if: the candidate term is not lower than its own term, and the server has not already voted for some other server for the current term.
This election mechanism is achieved with \I{RequestVote} requests and responses.

When a candidate for a given term is granted a majority of votes, it becomes the leader for this term and starts sending heartbeat messages to all other servers.
These heartbeat messages let other servers know about the new leader, and other candidates step down to the follower state.

A leader regularly sends heartbeat messages to reset the timeout counter of its followers, and thus prevent new elections.
Yet, if the leader server crashes or becomes isolated by a network partition, other servers stop receiving its messages and soon trigger a new leader election.

The timeout durations are chosen randomly within certain bounds, such that there is a high chance for an election to successfully establish a leader.
The frequency at which heartbeat messages are sent is chosen accordingly, to minimize the opportunities of spurious elections while a leader is alive and able to communicate with its followers.
The Raft paper \cite{Ongaro-Ousterhout-14} features a proper discussion on the choice of relevant timeout duration to use in practical implementations.

Once a leader is elected, it becomes responsible for log replication.

\subsection{Raft log replication}

A leader handles client commands by adding them to its own log, and making sure to replicate these log updates among its followers.
The log is made of \I{entries}, each entry containing a client command and the term of the leader that handled this command.

To achieve log replication, a leader exchanges \I{AppendEntries} requests and responses with its followers, to append one or more entries in their log.
An entry must be replicated by a majority of servers before it can be considered \I{committed}, i.e. ready to be applied to the service state machine.
Each server maintains a \I{commit index} that indicates the log index up to which entries are known to be committed.

Failures may trigger an election while the current leader still has some entries not yet replicated by a majority of servers.
The new leader for the next term may append new entries that do not match the ones that were not yet committed.
In such situations, the new leader must send AppendEntries requests that overwrite the tail of some followers log, effectively replacing the entries that are beyond the commit index by new ones that are imposed by the new leader.

When a leader is elected, it does not know the commit index of its followers.
It may try to replicate entries whose indexes are too far beyond some of the followers commit index.
In such cases, a follower can refuse an AppendEntries request: the leader then retries to replicate entries with a lower index until it catches up with the follower commit index.


\section{Modeling Raft in LNT}\label{MODEL}

Our LNT model is based on the Raft paper \cite{Ongaro-Ousterhout-14}, and the \TLA specification of the protocol \cite{Ongaro-14}.
Raft contains several optional features, such as dynamically changing the number of servers, or compacting the log.
Our model is restricted to the core of the protocol, i.e. leader election and log replication.
For the sake of being comparable, we tried to stay close to the variable names used in the \TLA specification.
After a primer on LNT, we describe our Raft model in a top-down fashion.

\subsection{LNT: a primer}\label{primer}

LNT is a formal language inspired by the LOTOS~\cite{ISO-8807} and E-LOTOS \cite{ISO-15437} process algebras.
Its syntax is close enough to regular programming languages that most of the language should be understandable to someone experienced in programming.
The detailed definition of the language is available in the LNT to LOTOS translator manual \cite{Champelovier-Clerc-Garavel-et-al-10-v6.7}.
This primer focuses on illustrating the interaction between concurrent processes.

\begin{figure}
    \centering
    \begin{subfigure}[t]{0.49\textwidth}
\begin{lstlisting}[]{Name}
module primer is

channel CalcOp is
  (nat, nat, nat)
end channel

process Calc [Add, Mul: CalcOp] is
  var op1, op2, res: nat in
    loop
      select
         Add(?op1, ?op2, ?res)
           where res == (op1 + op2)
      [] Mul(?op1, ?op2, ?res)
           where res == (op1 * op2)
      end select
    end loop
  end var
end process
\end{lstlisting}
    \end{subfigure}
    ~ 
    \begin{subfigure}[t]{0.49\textwidth}
\begin{lstlisting}[]{Name}
process User [Op: CalcOp] (a, b: nat) is
  var result: nat in
    Op(a, b, ?result)
  end var
end process

process Main [Add, Mul: CalcOp] is
  par Add, Mul in
     Calc[Add, Mul]
  || par
        User[Add](1, 2)
     || User[Mul](3, 4)
     end par
  end par
end process

end module
\end{lstlisting}
    \end{subfigure}

    \includegraphics[width=0.55\textwidth]{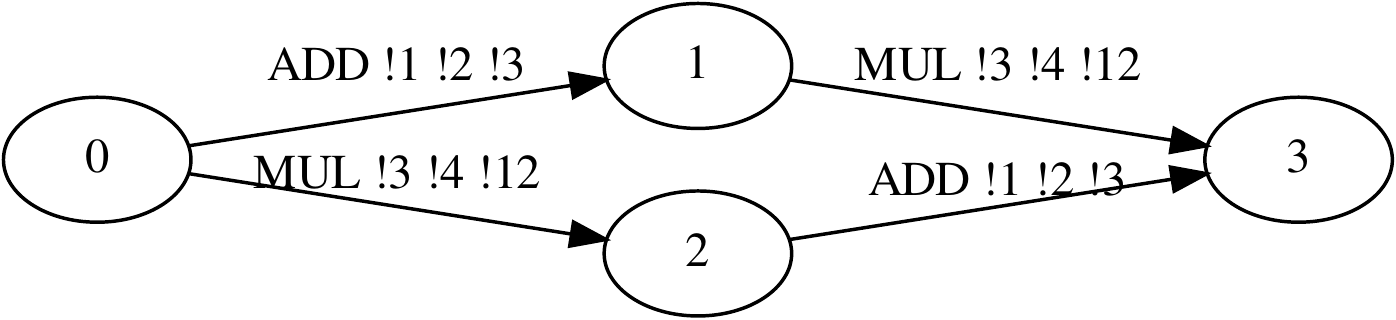}    

    \caption{Source code and state space of a small Calculator example of LNT}\label{fig:primer}
\end{figure}

In LNT, concurrent processes interact via an \I{action} on a \I{gate} to achieve both synchronization and data exchange.
Figure~\ref{fig:primer} illustrates LNT with a small example that defines three processes: \T{Calc} is a calculator, \T{User} is a user of this calculator, and \T{Main} is the top-level parallel composition of \T{Calc} and \T{User} processes.
\T{Calc} loops on offering actions on either gate \T{Add} or \T{Mul}, via the non-deterministic choice operator \T{select}.
Any action on one of these gates involves three natural numbers (built-in type \T{nat}), as imposed by the \T{CalcOp} \I{channel}.
For both calculator operations, the first two naturals are the operands and the third one is the result.
The \T{where} condition imposes a relation between the operands and the result, such that only the result corresponding to the operands is allowed in an action.

\T{User} takes one of the calculator operations as a gate argument (between square brackets), its two operands as data arguments (between parentheses), and does an action on that gate to perform the operation.
The value obtained in the \T{result} variable is imposed by the logic in \T{Calc}.

\T{Main} is the default process name for the model entry point.
It uses the \T{par} operator to instantiate a parallel composition of one \T{Calc} process and two \T{User} processes.
The inner \T{par} composition indicates that the two \T{User} processes run in parallel but do not synchronize on any gate.
However, the outer \T{par} composition indicates that any action on gates \T{Add} or \T{Mul} must synchronize \T{Calc} and one of the two \T{User} processes.
Another way to define which gates synchronize with which processes is to list gates before each process in a parallel composition.
For instance, our parallel composition can alternatively be written as:
\begin{lstlisting}[]{Name}
par
   Add, Mul -> Calc [Add, Mul]
|| Add -> User [Add] (1, 2)
|| Mul -> User [Mul] (3, 4)
end par
\end{lstlisting}

The graph illustrates the state space of this example, as generated by CADP tools: it is represented as a labeled transition system (LTS) with the actions as labels---no other information is retained.
We can see that both operations lead to relevant results, thanks to the \T{where} conditions.
More importantly, this LTS contains all possible execution paths: as both users run in parallel, there is no order imposed on which one of them interacts first with the calculator.
Hence, the LTS has a typical diamond shape that represents several possible execution paths leading to the same final state.

Now that we saw an overview of LNT, let's see how we use it to model a distributed system protocol like Raft, with a top-down approach.

\subsection{Raft model parameters}

The Raft protocol can be used between any number of servers, but our LNT model has to instantiate a specific Raft configuration among a fixed amount of servers.
In the same vein, we need to bound the number of terms, and the number of client interactions, otherwise the state space would be unbounded.
We define these three bounds as global configuration constants, which can be easily changed to obtain different configurations.

LNT does not have global constants \I{per se}.
Annex~\ref{a-params} illustrates how we can use functions with no arguments to define the \T{MaxTerm}, \T{MaxServerID} and \T{MaxClientInteraction} constants.
Server IDs range from one to \T{MaxServerID}, included: the zero ID is reserved as a ``nil'' value, which will prove useful in Section~\ref{message-reception}.
We also declare here the \T{NatArray} type, indexed by the server IDs, and the \T{Majority} function which returns the number of servers needed to get a majority.

\subsection{Top-level parallel composition}

In Annex~\ref{a-toplevel}, the \T{Main} process is our model entry point.
It defines the top-level parallel composition of three kinds of processes: \T{Network}, \T{Server} and \T{Clients}.
The nested parallel composition is such that: an action on gate \T{Send} or \T{Recv} involves the network and one of the servers, an action on \T{Client} involves a server and a client, and an action on \T{Crash} involves a server.
This enables servers to interact by sending and receiving messages to and from the network via \T{Send} and \T{Recv}, clients to interact with servers via \T{Client}, and server crashes to be monitored via \T{Crash}.
The possible data exchange on these gates is typed via the \T{ID} and \T{Comm} channels declared here.

Each server receives its ID as an argument.
The number of server instances in this parallel composition must match the \T{MaxServerID} constant.
Changing the number of servers is a matter of adding or removing server instances here, and updating \T{MaxServerID} and the range of \T{NatArray} accordingly.

\subsection{Interaction with clients}\label{interaction-with-clients}

The \T{Clients} process in annex~\ref{a-clients} is a simple abstraction of how clients may interact with the servers.
Neither the \TLA specification nor our model include the details of a service finite state machine, such that we do not need to model client commands, or server responses to clients.
Thus, client-server interaction boils down to an action on gate \T{Client}, where the only data is the ID of the leader with which the client interacts.
This is done in a loop which role is to bound the number of client interactions.

\subsection{Network communication}

Raft claims to be robust to the reordering, loss or duplication of messages.
Annex~\ref{a-network} contains our model for such an unreliable communication network.

We start by defining a message \T{Payload}: Raft messages always contain the sender's current term, together with the request or response of a \I{remote procedure call}\footnote{We use ``remote procedure call'' in this context to keep the naming coherent with the Raft paper.} (RPC), defined later.
A \T{NetworkMessage} encapsulates the origin and destination server ID, along with a payload.
A \T{NetworkMessageSet} is simply a set of network messages.

The \T{Network} process is responsible for transferring messages between servers, but it can also loose or duplicate messages.
It interacts with servers through the \T{Send} and \T{Recv} gates, which are named from a server point of view: the network actually obtains a message through an action on \T{Send}, and delivers it through an action on \T{Recv}.

This process is structured as an infinite loop on a non-deterministic choice of possible actions.
The network has two alternatives to obtain a message: in the first one, the message is successfully stored in the bag of messages; in the second one, the message is obtained but not stored: this models message loss.
There are also two alternatives for message delivery: in the first, the delivered message is effectively removed from the bag of messages; in the second, it is kept in the bag: this models message duplication.
In both cases of message delivery, any message stored in the bag may be transmitted: this models message reordering.

\subsection{Server}

Annex~\ref{a-server} presents the \T{Server} process, along with the declaration of a few types.
The \T{ServerState} type illustrates how to define an enumerated type with the possible states of a server.
The \T{Entry} type would in principle be made of both a term and a client command, yet as we do not model client commands (see section~\ref{interaction-with-clients}), only the term remains.
Both \T{EntryList} and \T{NatSet} types are pretty self-explanatory: they represent a list of entries, and a set of naturals, respectively.
The \T{LastTerm} function returns the term of the last entry in the server's log, or zero if this log is empty.

Raft servers may interact through two kinds of RPC, one for each of the protocol phases: the leader election is achieved by the \I{RequestVote} RPC, the log replication by the \I{AppendEntries} RPC.
The \T{RPC} type defines the request and response messages for these two RPC.
Note that the \T{matchIndex} field of \T{AppendEntriesResponse} is not mentioned in the Raft paper, but is present in the \TLA specification.

The \T{Server} process models the behavior of a Raft server.
After the initialization of its state variables, it follows a structure similar to the network process, i.e. an infinite loop on a non-deterministic choice between a few possible behaviors: hit a timeout, receive an RPC message, or interact with a client.
On top of that, the server may crash at any time.

\subsubsection{Timeout}\label{timeout}

When a leader times out, it sends AppendEntries requests to all its followers, whether to actually perform log replication or just as a heartbeat to prevent them to time out.
This is achieved by calling the \T{BroadcastAppendEntriesRequest} sub-process, which is defined underneath the \T{Server} process.

If a server times out while not being a leader, it becomes an election candidate for a new term.
As we bind our model on the number of terms servers may explore, we use the \T{stop} operator to disable progress on this \T{select} branch if our current term is already at the limit.
Note that this only stops the server from triggering a new election, it can still perform behaviors of the other \T{select} branches, like RPC message handling, within this term.

If the server's term is still within bounds to be able to trigger a new election, then the server increments its term, becomes candidate, votes for itself and sends a vote request to all the other servers.
In the \TLA specification, a candidate does not vote for itself, but rather sends itself a vote request.
We choose to have candidates vote for themselves because it is likely how a real implementation would behave, and it reduces the state space by having less vote requests being sent.

\subsubsection{Message reception}\label{message-reception}

A server receives a message from the network through an action on the \T{Recv} gate.
Here, we start by declaring the variables needed to unpack message contents, all named with a ``\T{m}'' prefix to mirror the \TLA specification.
Then, the action on gate \T{Recv} stores the message sender ID in \T{mFrom}, while the message payload is directly unpacked in \T{mTerm} and \T{mRPC}.

If the message is stale, i.e. the sender's term \T{mTerm} is smaller than our current term, then we do not bother handling it further.
This is a difference with the \TLA specification, which drops stale RPC responses, but does reply to stale RPC requests in order to let the sender quickly become aware of a new term.
This behavior makes sense to minimize the time window during which a server is using an outdated term.
Still, we can safely ignore stale requests, and here we do so to reduce the model state space.

If the message is not stale, then we check whether we should update our own term: this is the equivalent of the \TLA specification \T{UpdateTerm} part.
Updating to a higher term means falling back into the follower state, and also store zero as a ``nil'' server ID into \T{votedFor}, to indicate that we did not yet vote for anyone in this term.
This is where having zero as a ``nil'' ID proves to be useful.

Finally, we use the pattern matching operator \T{case} to conveniently dispatch the RPC message according to its type.
Each RPC message type has a corresponding handler, which receives the message contents extracted into variables via the matched pattern.
An argument must be prefixed by ``\T{!?}'' if it is an ``\T{in out}'' argument that may be modified by the handler.

We briefly describe the handlers, whose names mirror the ones in \TLA.
For a full discussion of why these behaviors safely implement a consensus, see the Raft original papers \cite{Ongaro-14, Ongaro-Ousterhout-14}.

\paragraph{RequestVoteRequest handler (annex~\ref{a-handlerequestvoterequest}).} 
The server checks whether to grant its vote or not based on the status of its log, and whether it has already voted for someone else or not.

\paragraph{RequestVoteResponse handler (annex~\ref{a-handlerequestvoteresponse}).} 
The server stores the vote response, and proceeds to become leader if it received a majority of votes.

\paragraph{AppendEntriesRequest handler (annex~\ref{a-handleappendentriesrequest}).}
The server checks whether its log is ready to receive the new entries.
If not, it declines the request.
Otherwise, it accepts the request and appends the entries, possibly by removing uncommitted entries at the tail of its log.
The server commit index is updated to the leader commit index, and a match index is computed to let the leader know up to which index this server logs is matching the leader's one.
While the AppendEntries RPC is designed to enable the update of several entries in one call, our model reflects the \TLA specification choice of restricting updates to one entry at a time.
Hence, an AppendEntries request is either a heartbeat message with zero entries, or a request to update a single entry, as verified with an assertion.

\paragraph{AppendEntriesResponse handler (annex~\ref{a-handleappendentriesresponse}).}
The server keeps track of its followers' current matching indexes in the \T{nextIndex} and \T{matchIndex} arrays, which are updated accordingly to the AppendEntries response.
Then, it tries to increase its commit index based on whether a majority of servers now agree on sharing the same entry at a higher log index.

\subsubsection{Client interaction}

The last alternative in the server main loop (back to annex~\ref{a-server}) is to interact with a client.
Only a leader may interact with clients, hence the action on gate \T{Client} is guarded by a check on the current server state.
Remember that client commands are not part of our model, so a client interaction results in a new entry containing only the leader's current term.

The AppendEntries requests related to this new log entry will be sent after a timeout (see section~\ref{timeout}).
One may expect a real implementation to send an AppendEntries request right after a client interaction, yet we know that at least in the case of Consul, the leader actually waits for its timeout, to be able to batch an update of all the client interactions that may happen during a leader timeout duration.
This choice may be detrimental to latency, but beneficial to throughput.

\subsubsection{Crash}

Servers may crash at any time.
The careful reader will have noticed that the whole main loop is encapsulated in a \T{disrupt} statement, which enables an action on gate \T{Crash} to happen at any point during the server execution.
One may think that the crash could simply be one of the \T{select} alternative, but this would not enable a crash to happen e.g. in the middle of a heartbeat broadcast.
This illustrates how convenient is the \T{disrupt} operator to model behaviors that may pertubate others at any point.






\subsection{Issues with the original \TLA specification}

Modeling Raft in LNT highlighted two issues with the original \TLA specification.

The first\footnote{\url{github.com/ongardie/raft.tla/blob/34cdd49d22615426ea00a6605b95be57b3cab49a/raft.tla\#L478}} is a minor typo: a ``\T{matchIndex'}'' which should have been a ``\T{matchIndex}'', without the apostrophe that denotes a variable's next state in \TLA.
This issue in \T{AdvanceCommitIndex} was minor and made no practical difference.

The second\footnote{\url{groups.google.com/g/raft-dev/c/yu-wOUx-gnA/m/VsM49xpFPwcJ}} issue is related to a missing server state change: candidates that received a heartbeat message from the leader who won the term election would not step down to become followers.
In practice, a candidate behaves almost like a follower, it just will not grant its vote to any other server, so this issue did not impact Raft manual proof.
Yet this proves that discrepancies between the intended behavior and the formal specification can appear by accident.


\section{Discussion on the modeling of distributed systems}\label{DISCUSSION}

While our model is specific to the Raft protocol, it invites some remarks on the
act of specifying distributed systems in general.
In this section, we discuss various aspects of our experience with developing the formal model of a distributed system.





\subsection{Iterative development with LNT and CADP}

Based on our experience with Raft, we argue that LNT and CADP offer a good environment to iteratively develop a formal specification of a distributed system.
With regular programming, the ability to experiment with a quick feedback loop can greatly improve the development experience.
In practice, this can be achieved with fast modify-compile-execute cycles, or by using interpreted languages which offer read-eval-print-loop interfaces.
We consider that writing a formal specification is close to writing a program, except that we operate with a formal language.
Having a quick iteration feedback loop is still relevant in a formal context.

With this in mind, the following features of LNT and CADP have been very helpful.
First, the LNT syntax is close to mainstream programming languages.
This cannot be said of numerous other formal languages, which roots in theoretical computer science often lead to an exotic syntax.
Second, LNT is a strongly typed language.
Type checking can catch many issues as early as compilation time.
Beside type errors, CADP is also able to statically report useful warnings about e.g. unused variables, or unreachable actions.
Not all formal languages are typed, and in particular \TLA is not typed.
Third, LNT features an \T{assert} statement, which will stop state-space generation with an error if violated.
This enables to quickly spot some bugs, without having to wait for a complete state-space generation.
Fourth, although not demonstrated here, LNT parallel composition enables to synchronize an arbitrary number of processes (two or more) on a single action.
Many languagues can only express process interaction between a pair of processes, such that any kind of barrier synchronization between more than two processes requires a protocol built on top of pair-wise interaction.
The ability to synchronize an arbitrary number of processes can prove very useful in some models \cite{Garavel-Serwe-17}, and lead to smaller state-space by factoring barrier synchronization protocols into a single action.

Fifth, CADP can generate state spaces of a given model in both implicit and explicit forms.
An implicit state space is effectively a dynamic library offering relevant primitives to generate the list of reachable states from a given state.
An explicit state space is the actual LTS containing all the states and transitions of a model, obtained by exhaustively exploring all reachable states from the model initial state.
Generating the explicit state space can take a significant time, which may hinder the iterative development process.
The implicit state space is typically generated within seconds, and can then be explored manually to check e.g. whether a given execution path is reachable.
This can prove very handy when trying modifications on the model being developed.



\subsection{Generic skeleton for distributed systems}

A distributed system can arguably be defined as \I{node} processes that interact via some sort of network.
In that sense, LNT models of distributed systems are very likely to have a top-level parallel composition similar to ours, i.e. a series of nodes (here, the Raft servers) communicating over a network.

Our \T{Network} process is already generic: it only knows about a message origin, destination, and payload.
What the payload actually contains can be easily adapted to the needs of other protocols.

While our \T{Server} process is specifically implementing Raft, we argue that its general structure can be used as a template for distributed system nodes.
This structure is expressed in the \T{GenericNodeSkeleton} process of annex~\ref{a-nodeskeleton}.
The node initializes its state, then enters a main loop where it reacts to the reception of messages, or other local events.
The \T{disrupt} operator lets us easily model a crash at any time.
Note that such generic structures have already been captured by established distributed system frameworks, like the \T{gen\_server}\footnote{\url{https://erlang.org/doc/design_principles/gen_server_concepts.html}} module in Erlang.

\subsection{A library of network models}

The genericity of our network process means that it can be easily replaced by other ones, with different communication behaviors.
As an example, annex~\ref{a-reliablenetwork} presents the \T{ReliableNetwork} process through which messages may be re-ordered, but neither lost nor duplicated.
This alternative network model is as simple to use as dropping it in place of the \T{Network} process in the top-level parallel composition.

This ability to swap network models brings up two major benefits.
First, using a reliable network helps in iteratively developing a model.
A reliable network will generate smaller state spaces, because its behavior is a subset of the unreliable one, which leads to less possible message interactions between nodes.
As an example, consider a Raft configuration of two servers, one term and one client interaction: with the unreliable \T{Network} process, this configuration leads to an LTS of 11,862,015 states and 74,821,042 transitions.
With the \T{ReliableNetwork} process, the state space is down to 22,311 states and 108,176 transitions, that is three and two orders of magnitude smaller, respectively.
Thus, a reliable network helps to get a quick development feedback loop by being able to generate smaller state spaces.
While those state spaces are not relevant for a proper verification, they can still be used as a proxy: if an issue is found with the reliable network, then it will appear with the unreliable network as well.

Second, being able to swap network models enables to test a system under a whole spectrum of network behaviors.
For instance, some network models could capture the guarantees offered by Unix sockets in TCP or UDP mode.
Others could model the communication guarantees of the run-time of some specific distributed programming languages.
We could then have a whole library of network models, such that the same distributed system can be easily verified against each of them.
This would help to gain confidence on which requirements are actually needed for a real implementation.
It could also help in making some existing distributed algorithms become robust to some communication failures (e.g. message duplication) that were not considered in its original design, by verifying the algorithm against a less reliable network model and then improving the algorithm until it supports this alternative network model.

\subsection{Keep the state space minimal}

When editing an LNT specification, it is recommended to try and keep the state space as small as possible, while still exploring all relevant execution paths.
This balance can be risky when, for the sake of minimizing the state space, the model restricts some execution paths that may well lead to some issues in the real implementation.
Here, we discuss a couple of techniques we used.

First, events surfaced as actions on gates should be kept to a minimum.
In the state space LTS, only the actions appear as transitions, so the less actions there are, the smaller the state space.
Still, there should be enough actions to analyze the state space and look for issues.
For distributed systems, a rule of thumb seems to be to surface events related to either inter-process communication, or outstanding local events.
In Raft, communication events happen on gates \T{Send} and \T{Recv}, while actions on gates \T{Client} and \T{Crash} are, as long as the protocol is concerned, local to a Raft server.
One could argue that actions on \T{Client} are not strictly necessary to surface, since we could infer whether a leader had a client interaction by observing its AppendEntries requests.
Still, we choose to model them as explicit actions because it makes it easier to reason about the resulting state space.

Second, some of the action interleavings may be avoided by forcing a specific order of execution.
For instance, in our Raft model a leader broadcasts AppendEntries requests in the order of server IDs (e.g. leader 1 would send requests to follower 2, and then follower 3).
If we wanted to be truly exhaustive, we should model this broadcast as sending requests in any order (e.g. leader 1 could send to either follower 2 or 3, and then to the other one).
Yet we argue that exploring all these sending orders is irrelevant here, because of the asynchronous nature of communications.
What matters is that the receiving order is not fixed, and in our case all possible receiving orders will be explored thanks to the network process which buffers messages and delivers them in any order.
Based on this reflection, we decided that it was acceptable to impose the sending order, and thus reduce the state space along the way.

This example demonstrates how it is sometimes acceptable to impose an execution order that reduces the state space, but one should be genuinely careful about deciding when it is safe to do so.
It is very easy to restrict the state space in a way that removes some of the possible execution paths that are actually problematic.
These decisions depend not only on the system being modeled, but also on the kinds of verification that are performed on the state space.
The obliteration of some execution paths may be irrelevant to a given verification property, but crucial to an other.
Therefore, great care should be taken when using this technique.










\subsection{Working around unreliable communication}

Could it be possible to safely reduce the state space by bypassing the modeling of unreliable communication altogether?
Many distributed system protocols or algorithms are designed to cope with unreliable network, using well-known techniques.
For instance, making messages idempotent is a classic way to cope with message duplication: receiving a message once or several times does not make a difference.
If we take for granted that nodes are robust to any message duplication, why bother modeling message duplication in our specification?
Removing message duplication from our model would help to reduce the state space by a significant amount.

In the case of our Raft model, messages are not only idempotent, but servers also discard any message that has a lower term than the server's current term.
Thus, besides removing message duplication from our network model, we could go further and make the network track the current term of each server and drop all buffered outdated messages that would be dropped by the destination server anyway.
This could reduce the state space by removing some \T{Recv} actions.

These considerations could pave the way to analyze state spaces of configurations with higher bounds on the number of servers, client interactions and terms, hence gaining more confidence in the verification.
Yet, it is also very easy to overlook how such bypassing may cut away corner-cases that are relevant to expose bugs that could happen in real implementations.
In other words, we are back to this subtle question: on which grounds can we be certain that it is safe to bypass some behaviors?
Such decisions would be best based on some formal verification or proof of specific well-known solutions, such that we can safely put those bypasses in place.
For instance, a proof that message duplication can be safely bypassed in modeling if all messages are idempotent would enable to use this technique in any distributed system model.








\subsection{Formal model and implementation: bridging the gap}

Several Raft implementations now exist, but how can we know whether they are correct?
Some tools specialize in testing distributed systems.
In recent years, the Jepsen\footnote{\url{https://jepsen.io}} tool has been used to analyze more than twenty distributed systems, finding issues in almost all of them, including systems that are based on Raft, like etcd.\footnote{See etcd analysis published on January 2020: \url{https://jepsen.io/analyses/etcd-3.4.3}}

More generally, the question is how to bridge the gap between the formal model of a protocol and its actual implementation.
\TLA does not have a code-generation feature.
Other projects, like Verdi~\cite{Wilcox-Woos-et-al-15}, are explicitly designed to generate code from the very same specification on which a proof is conducted.
As a matter of fact, Verdi has been used to model Raft and obtain an implementation directly from the model~\cite{Woos-Wilcox-et-al-16}.
In the model-checking world, the Distributed LNT Compiler~\cite{Evrard-16} enables the generation of a distributed implementation (in C) of an LNT model.
Yet, its protocol to implement synchronization on gates assumes a reliable message passing between processes, so it is not best suited to generate implementations of distributed systems.
Still, it would be possible to specialize this tool and have the network actions inthe model being replaced by actual call to e.g. Unix socket primitives.
Section~6 of \cite{Evrard-15} conducts a performance comparison between Consul and the code generated by DLC from an LNT Raft specification:
the DLC version can process 1000 client commands on a cluster of 3 servers in 2.3 seconds, where Consul requires 0.5 second.
DLC is slower than Consul in the general case, but Consul also features an optimization that buffers client commands on the leader to treat them in batch to favour throughput over latency, whereas the DLC version triggers an \I{AppendEntries} RPC round for each client command.
In any case, this study reports that DLC-generated implementations can achieve inter-LNT-process communication over TCP between separate machines in less than half a millisecond.


\section{Conclusion}\label{CONCLUSION}

We gave an overview of the Raft consensus protocol followed by a presentation of how we specified it in LNT.
Our model clearly separates the behavior of Raft servers on one side, and the behavior of the network on the other.
Having the network as a separate, generic entity enables not only to reuse this network model to be reused for other systems, but also to be replaced with other network behaviors.
This can help during the development of a model, to keep the state spaces from growing too big too early.
Our model was written to be comparable with the original \TLA specification of Raft, in which we found two issues.

We then conducted a discussion on the formal specification of distributed systems in general.
As opposed to proof systems, we argue that LNT and CADP offer an iterative development environment that provides a quick feedback loop.
We also discussed various techniques for safely reducing the state space of a model, while still capturing all the execution paths relevant for verification.

Our Raft model only contains the core of the protocol, and a possible future work would be to extend it to cover optional features, like dynamic configuration changes or log compaction.









\subsection*{Acknowledgements}

The author would like to thank the MARS workshop editors and reviewers
for their valuable feedback. Special thanks also goes to Hubert
Garavel and Frédéric Lang from the Inria CONVECS team for their
support in running some experiments.


\bibliographystyle{eptcs}
\bibliography{bibl-raft}

\newpage

\appendix

\lstset{
  language=LNT,
  basicstyle=\ttfamily\small,
  columns=fullflexible, 
  commentstyle=\rmfamily\itshape,
  keywordstyle=\rmfamily\bfseries,
  xleftmargin=\parindent
}

\section{The LNT specification of Raft}\label{SPECIFICATION}

\subsection{Global configuration parameters}\label{a-params}
\begin{lstlisting}
-- Global parameters to bound the system

-- Maximum term to be allowed
function MaxTerm: nat is
  return 3
end function

-- Maxium number of client interaction
function MaxClientInteraction: nat is
  return 3
end function

-- Server IDs are within 1..MaxServerID included
-- Zero is reserved as the "nil" server ID
function MaxServerID: nat is
  return 3
end function

-- The range of NatArray must be 1..MaxServerID
type NatArray is
  array[1 .. 3] of nat
end type

-- Majority returns the number of servers needed to get a majority
function Majority : nat is
   return (MaxServerID div 2) + 1
end function
\end{lstlisting}

\subsection{Top-level parallel composition}\label{a-toplevel}
\begin{lstlisting}
-- For action that only exchange a server ID
channel ID is
  (nat)
end channel

-- For general communication between servers
channel Comm is
  (nat, nat, Payload)
end channel

-- Entry point: parallel composition of Network, Servers and Client
process Main [Send, Recv: Comm, Client, Crash: ID] is
  par
     Send, Recv -> Network[Send, Recv]
  || Client -> Clients[Client]
  || Send, Recv, Client ->
       par
          -- The number of servers here must reflect MaxServerID
          Server[Send, Recv, Client, Crash](1)
       || Server[Send, Recv, Client, Crash](2)
       || Server[Send, Recv, Client, Crash](3)
       -- Additional servers can be added with e.g.:
       -- || Server[Send, Recv, Client, Crash](4)
       end par
  end par
end process
\end{lstlisting}

\subsection{Clients}\label{a-clients}
\begin{lstlisting}
-- Clients bounds the number of client interactions
process Clients [Client: ID] is
  var n: nat in
    for n := 0 while n < MaxClientInteraction by n := n + 1 loop
      Client(?any nat)
    end loop
  end var
end process
\end{lstlisting}

\subsection{Network}\label{a-network}
\begin{lstlisting}
-- Payload encapsulates the message contents
type Payload is
  Payload(mTerm: nat, rpc: RPC)
  with "get"
end type

-- NetworkMessage is a message Payload, plus origin and
-- destination which correspond to ServerID's
type NetworkMessage is
  NetworkMessage(orig, dest: nat, payload: Payload)
  with "get"
end type

type NetworkMessageSet is
  set of NetworkMessage
  with "element", "length", "remove"
end type

-- Network models an unreliable communication network.
-- Messages can be lost, duplicated, and reordered.
process Network [Send, Recv: Comm] is
  var
    n: nat,
    orig: nat,
    dest: nat,
    payload: Payload,
    msg: NetworkMessage,
    msgBag: NetworkMessageSet
  in
    msgBag := {}; -- start with no message in the bag
    loop
      select
        -- Obtain a message, and store it in the bag
        Send(?orig, ?dest, ?payload);
        msgBag := insert(NetworkMessage(orig, dest, payload), msgBag)
      []
        -- Message loss: obtain a message, but do not store it
        Send(?any nat, ?any nat, ?any Payload)
      []
        -- Transmit any message from the bag, then remove it from the bag
        -- Note: by picking up any message from the bag, we model reordering
        n := any nat where (n > 0) and (n <= length(msgBag));
        msg := element(msgBag, n);
        Recv(msg.orig, msg.dest, msg.payload);
        msgBag := remove(msg, msgBag)
      []
        -- Message duplication: transmit it, but do not remove it from the bag
        n := any nat where (n > 0) and (n <= length(msgBag));
        msg := element(msgBag, n);
        Recv(msg.orig, msg.dest, msg.payload)
      end select
    end loop
  end var
end process
\end{lstlisting}

\subsection{Server}\label{a-server}
\begin{lstlisting}
type ServerState is
  Follower,
  Candidate,
  Leader
  with "==", "!="
end type

-- Entry content is restricted to a term, we do not model client commands
type Entry is
  Entry(term: nat)
  with "get"
end type

type EntryList is
   list of Entry
   with "element", "append", "empty", "length"
end type

type NatSet is
   set of nat
   with "length"
end type

-- RPC models the remote procedure calls of Raft
-- Each of RequestVote and AppendEntries has a request and a response message
type RPC is
  RequestVoteRequest(lastLogIndex, lastLogTerm: nat),
  RequestVoteResponse(voteGranted: bool),
  AppendEntriesRequest(prevLogIndex, prevLogTerm: nat, entries: EntryList,
                         leaderCommit: nat),
  AppendEntriesResponse(success: bool, matchIndex: nat)
end type

-- LastTerm returns the term of the last log entry
function LastTerm(log: EntryList) : nat is
  case log in
    {} -> return 0
  | any -> return get_term(element(log, length(log)))
  end case
end function

-- Server models the behavior of a Raft server
process Server [Send, Recv: Comm, Client, Crash: ID] (self: nat) is
  var
    -- server internal state
    state: ServerState,
    currentTerm: nat,
    log: EntryList,
    commitIndex: nat,
    -- related to RequestVote RPC
    votedFor: nat,
    votesGranted: NatSet,
    -- related to AppendEntries RPC
    nextIndex: NatArray,
    matchIndex: NatArray
  in

    -- This disrupt wraps the whole main loop to model a Crash at any time
    -- See the actual Crash action at the bottom of the process
    disrupt

      -- state initialization
      state := Follower;
      currentTerm := 0;
      log := {};
      commitIndex := 0;
      votedFor := 0;
      votesGranted := {};
      nextIndex := NatArray(0);
      matchIndex := NatArray(0);

      -- main loop
      loop
        select

          -- Timeout
          if state == Leader then
            BroadcastAppendEntriesRequest[Send](self, currentTerm,
                                                    commitIndex, log, nextIndex)
          else
            if currentTerm >= MaxTerm then
              -- Do not progress further
              stop
            else
              -- Trigger election for a new term
              currentTerm := currentTerm + 1;
              state := Candidate;
              -- Diff with TLA+ spec: we do vote for ourselves directly
              votedFor := self;
              votesGranted := {self};
              -- Broadcast vote requests
              var n: nat in
                for n := 1 while n <= MaxServerID by n := n + 1 loop
                  if (n != self) then
                    Send(self, n,
                          Payload(currentTerm,
                                   RequestVoteRequest(length(log),
                                                        LastTerm(log))))
                  end if
                end loop
              end var
            end if
          end if

        []

          -- Receive message
          var
            -- Variables to unpack message content
            mFrom: nat,
            mTerm: nat,
            mRPC: RPC,
            -- RequestVote
            mLastLogIndex: nat,
            mLastLogTerm: nat,
            mVoteGranted: bool,
            -- AppendEntries
            mPrevLogIndex: nat,
            mPrevLogTerm: nat,
            mEntries: EntryList,
            mLeaderCommit: nat,
            mSuccess: bool,
            mMatchIndex: nat
          in

            -- Network interaction to receive a message
            Recv(?mFrom, self, ?Payload(mTerm, mRPC));

            if mTerm < currentTerm then
              -- Drop stale message
              -- Diff with TLA+ spec: drop any message, where in TLA+
              -- the server would still reply to stale RPC *requests*
              null
            else

              -- This is the "UpdateTerm" part of the TLA+ spec
              if mTerm > currentTerm then
                currentTerm := mTerm;
                state := Follower;
                votedFor := 0 -- Value zero is the nil server ID
              end if;

              -- Dispatch to the relevant handler using pattern matching
              case mRPC in
                RequestVoteRequest(mLastLogIndex, mLastLogTerm) ->
                  HandleRequestVoteRequest[Send](self, currentTerm, LastTerm(log),
                                                     mLastLogTerm, length(log),
                                                     mLastLogIndex, mFrom, !?votedFor)
              | RequestVoteResponse(mVoteGranted) ->
                  eval HandleRequestVoteResponse(log, !?nextIndex, !?matchIndex,
                                                    mFrom, !?state, mVoteGranted,
                                                    !?votesGranted)
              | AppendEntriesRequest(mPrevLogIndex, mPrevLogTerm,
                                       mEntries, mLeaderCommit) ->
                  HandleAppendEntriesRequest[Send](self, mFrom, currentTerm,
                                                       mPrevLogIndex, mPrevLogTerm,
                                                       mEntries, mLeaderCommit,
                                                       !?commitIndex, !?log)
              | AppendEntriesResponse(mSuccess, mMatchIndex) ->
                  if state == Leader then
                    eval HandleAppendEntriesResponse(currentTerm, mFrom, mSuccess,
                                                        mMatchIndex, log, !?nextIndex,
                                                        !?matchIndex, !?commitIndex)
                  end if
              end case
            end if
          end var

        []

          -- Client interaction
          Client(self) where state == Leader;
          log := append(Entry(currentTerm), log)

        end select
      end loop

    by
      -- Crash may happen at any time
      Crash(self)
    end disrupt

  end var
end process

-- Helper process to broadcast AppendEntries requests
process BroadcastAppendEntriesRequest [Send: Comm] (self: nat,
                                                         currentTerm: nat,
                                                         commitIndex: nat,
                                                         log: EntryList,
                                                         nextIndex: NatArray) is
  var
    n, prevLogIndex, prevLogTerm, lastEntry : nat,
    entries: EntryList
  in
    for n := 1 while n <= MaxServerID by n := n + 1 loop
      if (n != self) then
        prevLogIndex := nextIndex[n] - 1;
        if prevLogIndex > 0 then
          prevLogTerm := get_term(element(log, prevLogIndex))
        else
          prevLogTerm := 0
        end if;
        lastEntry := min(length(log), nextIndex[n]);
        entries := SubSeq(log, nextIndex[n], lastEntry);
        Send(self, n,
             Payload(currentTerm,
                      AppendEntriesRequest(prevLogIndex, prevLogTerm,
                                             entries, commitIndex)))
      end if
    end loop
  end var
end process
\end{lstlisting}

\subsection{RPC handlers}

\subsubsection{HandleRequestVoteRequest}\label{a-handlerequestvoterequest}
\begin{lstlisting}
process HandleRequestVoteRequest [Send: Comm] (in self: nat,
                                                   in currentTerm: nat,
                                                   in lastLogTerm: nat,
                                                   in mLastLogTerm: nat,
                                                   in lengthLog: nat,
                                                   in mLastLogIndex: nat,
                                                   in mFrom: nat,
                                                   in out votedFor: nat) is
  var logOK: bool, voteGranted: bool in

    logOK := (mLastLogTerm > lastLogTerm) or
                ((mLastLogTerm == LastLogTerm) and (mLastLogIndex >= lengthLog));

    voteGranted := logOK and ((votedFor == 0) or (votedFor == mFrom));

    if voteGranted then
      votedFor := mFrom
    end if;

   Send(self, mFrom, Payload(currentTerm, RequestVoteResponse(voteGranted)))

  end var
end process
\end{lstlisting}

\subsubsection{HandleRequestVoteResponse}\label{a-handlerequestvoteresponse}
\begin{lstlisting}
function HandleRequestVoteResponse (log: EntryList,
                                      in out nextIndex: NatArray,
                                      in out matchIndex: NatArray,
                                      voter: nat,
                                      in out state: ServerState,
                                      mVoteGranted: bool,
                                      in out votesGranted: NatSet) is
  if (state == Candidate) and mVoteGranted then
    votesGranted := insert(voter, votesGranted);

    if length(votesGranted) >= Majority then
       -- This is the "BecomeLeader" part of the TLA+ spec
       state := leader;
       nextIndex := NatArray(length(log) + 1);
       matchIndex := NatArray(0)
    else
      use nextIndex;
      use matchIndex
    end if
  end if
end function

--------------------------------------------------
-- Helpers

function SubSeq (e: EntryList, in var i: nat, in var j: nat): EntryList is
   var n: nat, result: EntryList in
      result := {};
      -- restrict indexes
      if i == 0 then
         i := 1
      end if;
      j := min(j, length(e));
      for n := i while n <= j by n := n + 1 loop
        result := append(element(e, n), result)
     end loop;
     return result
   end var
end function
\end{lstlisting}

\subsubsection{HandleAppendEntriesRequest}\label{a-handleappendentriesrequest}
\begin{lstlisting}
process HandleAppendEntriesRequest [Send: Comm] (self: nat,
                                                     leaderId: nat,
                                                    currentTerm: nat,
                                                    mPrevLogIndex: nat,
                                                    mPrevLogTerm: nat,
                                                    entries: EntryList,
                                                    mLeaderCommit: nat,
                                                    in out commitIndex: nat,
                                                    in out log: EntryList) is
  var
    logOK : bool,
    index, matchIndex : nat,
    entry: Entry
  in

    logOK := (mPrevLogIndex == 0) or
                ((mPrevLogIndex > 0)
                  and (mPrevLogIndex <= length(log))
                  and (mPrevLogTerm == get_term(element(log, mPrevLogIndex))));

    -- Diff with TLA+ spec: we don't need to test if the request
    -- was stale (outdated term), because we drop any stale message
    -- before dispatching to handlers

    if not(logOK) then
      -- Decline request
      Send(self, leaderId,
            Payload(currentTerm, AppendEntriesResponse(False, 0 of nat)));
      -- commitIndex is an "in out" variable, we must use it in all execution paths
      use commitIndex
    else
      -- Accept request
      index := mPrevLogIndex + 1;
      if empty(entries) then
         -- Requests with no entry are heartbeat messages
         matchIndex := mPrevLogIndex
      else
        -- We mirror the TLA+ spec logic which limits to 1 entry at a time
        assert(length(entries) == 1);
        entry := element(entries, 1);
        if (length(log) >= index) and (get_term(element(log, index)) != entry.term) then
           -- Conflict, remove our log tail
           log := SubSeq(log, 1, index - 1)
        end if;
        if length(log) == mPrevLogIndex then
          -- Entry not yet in the log, append it
          log := append(entry, log)
        end if;
        matchIndex := mPrevLogIndex + 1
      end if;
      commitIndex := mLeaderCommit;
      Send(self, leaderId,
            Payload(currentTerm, AppendEntriesResponse(True, matchIndex)))
    end if
  end var
end process
\end{lstlisting}

\subsubsection{HandleAppendEntriesResponse}\label{a-handleappendentriesresponse}
\begin{lstlisting}
function HandleAppendEntriesResponse (currentTerm: nat,
                                         mFrom: nat,
                                         success: bool,
                                         mMatchIndex: nat,
                                         log: EntryList,
                                         in out nextIndex: NatArray,
                                         in out matchIndex: NatArray,
                                         in out commitIndex: nat) is
  if success then
    nextIndex[mFrom] := mMatchIndex + 1;
    matchIndex[mFrom] := mMatchIndex
  else
    nextIndex[mFrom] := max(1 of nat, nextIndex[mFrom])
  end if;

  eval AdvanceCommitIndex(currentTerm, log, matchIndex, !?commitIndex)
end function

--------------------------------------------------
-- Helpers

-- MajorityAgree returns true if a majority of servers have the same
-- entry at the given index
function MajorityAgree (index: nat, matchIndex: NatArray): bool is
  var n, nbagree: nat in
    -- start nbagree at 1 since the current leader always agree,
    -- but its own matchIndex is zero
    nbagree := 1;
    for n := 1 while n <= MaxServerID by n := n + 1 loop
      if index <= matchIndex[n] then
        nbagree := nbagree + 1
      end if
    end loop;
    return nbagree >= Majority
  end var
end function

-- MaxAgreeIndex returns the higher index for which servers agree
function MaxAgreeIndex (lengthLog: nat, matchIndex: NatArray): nat is
  var index: nat in
    index := lengthLog;
    while not(MajorityAgree(index, matchIndex)) loop
      index := index - 1
    end loop;
    return index
  end var
end function

function AdvanceCommitIndex (currentTerm: nat,
                             log: EntryList,
                             matchIndex: NatArray,
                             in out commitIndex: nat) is
  var index: nat in
    index := MaxAgreeIndex(length(log), matchIndex);
    if (index > 0) and (get_term(element(log, index)) == currentTerm) then
      commitIndex := index
    else
      use commitIndex
    end if
  end var
end function
\end{lstlisting}

\section{General distributed system modeling}\label{a-distribmodel}

\subsection{Node skeleton}\label{a-nodeskeleton}
\begin{lstlisting}
process GenericNodeSkeleton [Send, Recv, Crash, Other, ...] (...) is
  var
    ... -- Declare node state variables
  in
    disrupt
      ... -- Initialize state variables
      loop
        select
          -- Handle message reception
          Recv(...);
          ...
          Send(...) -- May trigger sending of other messages
        []
          -- React on other events: timeouts, local sensor reading, etc
          Other(...);
          ...
          Send(...)
        []
          ...
        end select
      end loop
    by
      Crash() -- Crash at any time
    end disrupt
  end var
end process
\end{lstlisting}

\subsection{Reliable Network}\label{a-reliablenetwork}
\begin{lstlisting}
process ReliableNetwork [Send, Recv: Comm] is
  var
    n: nat,
    orig: nat,
    dest: nat,
    payload: Payload,
    msg: NetworkMessage,
    msgBag: NetworkMessageSet
  in
    msgBag := {};
    loop
      select
        -- Obtain a message, and store it in the bag
        Send(?orig, ?dest, ?payload);
        msgBag := insert(NetworkMessage(orig, dest, payload), msgBag)
      []
       -- Transmit a message from the bag, in no particular order
        n := any nat where (n > 0) and (n <= length(msgBag));
        msg := element(msgBag, n);
        Recv(msg.orig, msg.dest, msg.payload);
        msgBag := remove(msg, msgBag)
      end select
    end loop
  end var
end process
\end{lstlisting}

\end{document}